\documentstyle[12pt]{article}
\begin{document}
\title{ANALYSIS AND CHARACTERIZATION OF COMPLEX SPATIO-TEMPORAL
PATTERNS IN NONLINEAR REACTION-DIFFUSION SYSTEMS}
\author{Nita Parekh, V. Ravi Kumar, and B.D. Kulkarni* \\
Chemical Engineering Division, \\ National Chemical Laboratory, \\
Pune - 411 008, INDIA}
\date{}
\baselineskip20pt
\parskip20pt
\maketitle
\vspace{2cm}
\begin{abstract}
Two important classes of spatio-temporal patterns, namely, spatio-temporal
chaos and self-replicating patterns, for a representative three variable
autocatalytic reaction mechanism coupled with diffusion has been studied.
The characterization of these patterns has been carried out in terms of
Lyapunov exponents and dimension density. The results show a linear
scaling as a function of sub-system size for the Lyapunov dimension and
entropy while the Lyapunov dimension density was found to rapidly
saturate. The possibility of synchronizing the spatio-temporal dynamics by
analyzing the conditional Lyapunov exponents of sub-systems was also
observed.
\end{abstract}
\pagebreak

\section{Introduction}

The study of temporal chaos -- based on the analysis of data from physical
measurements at a point in phase-space or using model equations defined as
a set of ordinary differential equations (ODE's) -- has matured
considerably. Thus, methods for characterizing the basic properties of
low-dimensional nonlinear systems are now reasonably well developed [1,2].
In comparison, the understanding of spatio-temporal chaos is at a less
developed level and presently considerable attention is being focussed on
this topic [1,2,3]. Examples of systems exhibiting complex spatio-temporal
patterns include hydrodynamic systems [4,5], thermal convection in fluids
[6], nonlinear optics [7], chemical reactions [8,9], excitable biological
systems [10], crystallization and solidification fronts [11], etc. The
phenomenological models in the form of nonlinear partial differential
equations (PDEs), e.g., the Swift-Hohenberg equation and its variants, the
Kuramoto-Sivashinsky equation, Ginzburg-Landau equation and
reaction-diffusion models have been developed to study the nonlinear and
nonequilibrium properties of these systems [2]. Other modelling strategies
such as Coupled Map Lattices (CMLs) [12], Cellular Automata (CA) [13],
lattice gas models and their derivatives [14] have also been used for
understanding the dynamical behavior of the spatially-extended systems.
These methods have been successfully employed for analyzing multi-phase
flow systems, reaction-diffusion behavior, polymeric dynamics, etc.

For the phase-space reconstruction and model building of the
spatio-temporal systems, attempts have been made to generalize the known
methods of analyzing temporal systems from time-series data [1]. For
higher dimensional systems, the analysis turns out to be computationally
very demanding because of the large number of spatial degrees of freedom
[15-18]. In this paper, we aim at exemplifying some features of sub-system
behavior which may help in the characterization of the spatio-temporal
behavior of reaction-diffusion systems exhibiting complex dynamics. We
shall focus attention on quantitatively estimating the degree of
space-time coherence in the apparently complex dynamics (e.g.,
spatio-temporal chaos and self-replicating patterns) that arise in a
representative nonlinear autocatalytic reaction model. Our analysis gives
interesting relationships in the system invariant properties as a function
of sub-system size. We have also attempted to study the dynamical
synchronization properties of this extended system from sub-system
analysis. The possibility of synchronizing the behavior of spatio-temporal
dynamics may have considerable implications in the control of the system
behavior.

\section{Model Development}

The objectives outlined above were studied by considering a specific
example from a fairly general class of mechanisms in chemically reacting
systems and commonly referred to in the literature as an autocatalator
[19,20]. The reaction scheme considered here is the three-step parallel
autocatalytic reaction mechanism [20] with competing interactions between
the chemical species $A$, $D$, and $B$ and expressed as,
\begin{eqnarray*}
A + 2B &\longrightarrow & 3B , \hspace {20pt} -r_1 = k_1 C_A C^2_B ,
\hspace {40pt} (I) \\ 
D + 2B &\longrightarrow & 3B , \hspace {20pt} -r_2 = k_2 C_D C^2_B ,
\hspace {40pt} (II) \\ 
B &\longrightarrow & C , \hspace {35pt} r_3 = k_3 C_B , \hspace {57pt} (III)
\end{eqnarray*}\noindent
with the rate expression $r_i$, $i = 1, 2, 3$ for each step given
alongside. As may be seen nonlinear feedback occurs due to the rates of
formation of species $B$ being autocatalytic in steps I and II while in
III the effect is counteractive and inhibitory. For a continuous flow
well-mixed cell, the model description for the above reaction mechanism
takes the following dimensionless form
\begin{eqnarray}
{dX_1\over dt} & = & 1 - X_1 - Da_1 X_1 X_3^2 \nonumber \\
{dX_2\over dt} & = & \beta - X_2 - Da_2 X_2 X_3^2 \nonumber \\
{dX_3\over dt} & = & 1 - (1 + Da_3)X_3 + \alpha (Da_1 X_1 + Da_2 X_2) X_3^2 ,
\end{eqnarray}\noindent 
where $X_i$, $i = 1, 2, 3$, respectively, represent the dimensionless
species concentrations of $A$, $D$, and $B$ relative to their
concentrations at the inlet to the cell from the surroundings. Here,
$Da_i$'s are dimensionless kinetic parameters associated with the reaction
steps (I-III) and the parameters $\alpha$ and $\beta$ are the feed
concentration ratios of species $B$ and $D$ with respect to $A$. An
analysis of the bifurcation map and dynamics of this cell is known to
exhibit features such as multistationarity, oscillations and low
dimensional chaos in system parameter space of $\alpha$, $\beta$ and
$Da_i$'s [20]. The corresponding model for a diffusion mechanism operating
in one spatial dimension $x$ may be written as
\begin{eqnarray}
{\partial X_1(x,t)\over \partial t} = &1& - X_1(x,t) 
- Da_1 X_1(x,t) X_3^2(x,t) + d_1 {\partial ^2 X_1(x,t)\over \partial x^2} 
\nonumber \\
{\partial X_2(x,t)\over \partial t} = &\beta& - X_2(x,t) 
- Da_2 X_2(x,t) X_3^2(x,t) + d_2 {\partial ^2 X_2(x,t)\over \partial x^2} 
\nonumber \\
{\partial X_3(x,t)\over \partial t} = &1& - (1 + Da_3)X_3(x,t) 
+ \alpha [Da_1 X_1(x,t) + Da_2 X_2(x,t)] X_3^2(x,t) \nonumber \\
&+& d_3 {\partial ^2 X_3(x,t) \over \partial x^2} ,
\end{eqnarray} \noindent 
where $d_i$, $i = 1, 2, 3$ are the diffusion coefficients of the species
$A$, $D$, and $B$ respectively. The Euler discretization of the Laplacian
yields
\begin{eqnarray}
{\partial X_1(j,t)\over \partial t} = & 1 & - X_1(j,t) 
- Da_1 X_1(j,t) X_3^2(j,t) \nonumber \\
& + & D_1 {[X_1(j+1,t) - 2X_1(j,t) + X_1(j-1,t)]} \nonumber \\
{\partial X_2(j,t)\over \partial t} = & \beta & - X_2(j,t) 
- Da_2 X_2(j,t) X_3^2(j,t) \nonumber \\
& + & D_2 {[X_2(j+1,t) - 2X_2(j,t) + X_2(j-1,t)]} \nonumber \\
{\partial X_3(j,t)\over \partial t} = & 1 & - (1 + Da_3)X_3(j,t) 
+ \alpha [Da_1 X_1(j,t) + Da_2 X_2(j,t)] X_3^2(j,t) \nonumber \\
& + & D_3 {[X_3(j+1,t) - 2X_3(j,t) + X_3(j-1,t)]} ,
\end{eqnarray}\noindent 
$D_i = d_i/(\Delta x)^2$; $\Delta x$ is the spatial mesh size of the
discretized lattice; and $j = 1, 2,\dots N$. This discretized model may be
considered to describe the dynamics of $N$ autocatalytic cells defined by
(1) and coupled through diffusion. The effects of diffusion may result in
an interplay between the local cell dynamics and this in turn may lead to
pattern formation. The number of degrees of freedom is now significantly
increased to $3N$ and the characterization of the system dynamics is not a
trivial task. Here, the diffusion coefficients of species $A$ and $D$ are
assumed equal and much greater than for species $B$, i.e., $D_1 = D_2 >
D_3$. This is in accordance with Turing's conjecture for the occurrence of
spatio-temporal patterns in biological systems [21]. It has been shown
that for appropriate choices of diffusion rates, a variety of stationary
spatio-temporal patterns such as target, striped, or hexagonal patterns
and travelling waves may develop in simple reaction-diffusion systems
[22,23]. Recently, this conjecture has been confirmed by experiments with
thin two-dimensional gel laboratory reactors uncontaminated by convection
effects [24,25].

For the first part of the study all the cells of the system were assigned
identical initial conditions and the parameter values were chosen
corresponding to single cell (1) exhibiting chaotic dynamics. The chaotic
nature of the single cell dynamics was confirmed by the positive value of
the maximum Lyapunov exponent for its temporal dynamics, $\lambda_{max}
\sim 1.36$. The extended system (3) dynamics was then simulated using the
fourth order Runge-Kutta algorithm with time step $\Delta t = 0.0002$. The
studies were carried out on a one-dimensional lattice with $N = 64$ and
periodic boundary conditions were given. In fig. 1, for time $t < 40$ in
dimensionless units, the system dynamics is seen to be spatially
correlated though temporally it is uncorrelated and chaotic. Note that the
diffusion mechanism is not active in this region because of identical
initial conditions assumed over the entire spatial domain, i.e., all the
cells evolve in phase with each other. Now at time $t = 40$, a
perturbation was given to the chosen eleven cells lying in the central
region of the lattice. This perturbation assumed that the dynamics of the
chosen cells progressively went out of phase with each other by time $t =
0.002$. A spread of this perturbation to the system boundaries by
diffusion, with a simultaneous loss in the spatial correlation, is
qualitatively seen in fig. 1. This loss in the spatial correlation was
seen to persist even at $t = 10,000$ time units (results not presented
here).

In the second part of this study we considered the system in the parameter
range where it might exhibit coherent patterns and one such interesting
pattern is shown in fig. 2. In this case, the perturbation was seen to
develop into peaks that self-replicate structurally in time. Similar
patterns have been observed in a two-variable model system [26,27]. Recent
experiments have also shown their presence unambiguously [25]. This
self-replicating behavior, is in a sense, analogous to what is observed
during the phenomenon of self-replicating growth of biological cells [28],
DNA and RNA [29,30], micelles [31], etc. For comparative purposes, it may
be noted that for the study in fig.2, the mathematical model for the three
step autocatalytic reaction mechanism, assumed the form similar to that
considered in [27] by suitable dimensionalization. i.e.,
\begin{eqnarray}
{\partial X_1(x,t)\over \partial t} = & 1& -  X_1(x,t) 
- Da_1 X_1(x,t) X_3^2(x,t) + d_1 {\partial ^2 X_1(x,t)\over \partial x^2} 
\nonumber \\
{\partial X_2(x,t)\over \partial t} = & \beta & - X_2(x,t) 
- Da_2 X_2(x,t) X_3^2(x,t) + d_2 {\partial ^2 X_2(x,t)\over \partial x^2}
\nonumber \\
{\partial X_3(x,t)\over \partial t} = & - & (1 + Da_3)X_3(x,t) 
+ \alpha [Da_1 X_1(x,t) + Da_2 X_2(x,t)] X_3^2(x,t) 
\nonumber \\
& + & d_3 {\partial ^2 X_3(x,t)\over \partial x^2} ,
\end{eqnarray}
We shall study the dynamical characterization of the above two widely
different and important class of spatio-temporal patterns, namely, (a) the
chaotic behavior and (b) the coherent self-replicating behavior.

\section{Characterization of Spatio-Temporal Dynamics}

We shall first discuss the characterization of chaotic dynamics exhibited
by this spatially extended system in terms of the system Lyapunov
exponents [32]. For the present system with three species interacting on a
one-dimensional lattice of size $N$ there exists $3N$ Lyapunov exponents
and their calculation is computationally very demanding. To alleviate this
problem we propose to analyze the Lyapunov spectrum of its sub-systems of
size $n_s (< N)$ as $n_s \rightarrow N$ and look for any interesting
relationships that may help in characterizing the system dynamics. We
shall now briefly discuss below the calculation of sub-system Lyapunov
exponents.

For the extended system (3) with $\kappa$ denoting the set of fixed
parameters, the temporal evolution of the concentrations of each of the
species for chosen values of $D_i$ may be functionally written as
\begin{equation}
{dX_i(k,t) \over dt} = F_{i,k} (X_i(k,t), X_i(k \pm 1,t), D_i,{\kappa}) ,
\end{equation} \noindent 
where the first index $i = 1,2,3$ denotes the respective species and the
second index $k = 1,\dots,n_s$ denotes a sub-system lattice site. For a
sub-system of size $n_s$, there are $3n_s$ independent variables and
therefore $3n_s$ exponents. These may be calculated by monitoring the
growth rate of $3n_s$ sets of orthonormal vectors $\delta X_i (k,t)$, $i =
1, 2, 3$, $k = 1,2,\dots,n_s$, in a linearized region of (3). The complete
set of linearized equations built around a reference system state ${\bf
X}_r (t) \equiv X_i (k,t)$ may then be written as
\begin{equation}
{d \delta {\bf X}(t) \over dt} = {\bf J} \delta {\bf X} (t) .
\end{equation}\noindent 
Here, $\bf J$ is the augmented Jacobian of size $3n_s \times 3n_s$
evaluated at ${\bf X}_r (t)$ and $\delta {\bf X} (t)$ refers to the
infinitesimal perturbation from this reference state. Using the
fundamental matrix $\phi (t,t_0)$ the $3n_s$ solutions of (6) may be
expressed in a general form as
\begin{equation}
\delta {\bf X} (t) = \phi(t,t_0) \delta {\bf X} (t_0) .
\end{equation}\noindent 
The above relation is a linear map of different vector spaces. That is,
$\phi (t,t_0)$ is the mapping of $\delta {\bf X} (t_0)$ (related to the
tangent space $E_0$ at the phase-space point ${\bf X}_r (t_0) = X_i
(k,t_0)$) to $\delta {\bf X} (t)$ (associated with the tangent space $E_t$
for the phase-space point ${\bf X}_r (t) = X_i (k,t)$) with
\begin{equation}
\phi(t,t_0) = \phi(t,t_{n-1}) \dots \phi(t_2,t_1) \phi(t_1,t_0) .
\end{equation}\noindent 
The time averaging of $\delta {\bf X} (t)$, along with Gram-Schmidt
orthonormalization at periodic intervals, yields the $3n_s$ sub-system
Lyapunov exponents as
\begin{eqnarray}
\lambda^{(s)}_j = \lim_{t \rightarrow \infty} sup {1 \over t} 
ln { {\mid \delta {\bf X} (t) \mid} \over {\mid \delta {\bf X} (t_0) \mid} }, 
\hspace {30pt} j = 1 \dots 3n_s ,
\end{eqnarray}\noindent 
in a decreasing order. While calculating these exponents for the discrete
set of equations (3), fixed boundary conditions were assumed at the
sub-system boundary sites, i.e., at $k = 1$ and $k = n_s$ . The fixed
boundary conditions suggest that only for the purposes of evaluating the
$\lambda_j$, the flow of information from the outer site of the sub-system
boundaries, may be likened, to presence of noise in the analysis.

Using the above formalism we found that convergence of the sub-system
Lyapunov exponents was robust, although, for large $n_s$ the computer time
required was extremely high. In figs. 3a and 3b is shown the channelled
dynamics, considered from the central region in fig. 1, for sub-systems
$n_s = 7$ and $n_s = 31$, respectively. The convergence behavior of the
maximum sub-system exponent for these cases is presented in figs. 3c and
3d.

For the analysis of the complete system dynamics we consider the Kaplan
and Yorke conjecture [32,33] to calculate the effective sub-system
Lyapunov dimension $d^{(s)}_L$ defined as
\begin{equation}
d^{(s)}_L = j + {1 \over {\mid \lambda^{(s)}_{j+1} \mid }} \sum_{i=1}^j
{\lambda^{(s)}_i} ,
\end{equation}\noindent 
where $j$ is the largest integer for which the sum of the exponents,
$\lambda^{(s)}_1 + \dots + \lambda^{(s)}_j \geq 0$. In fig. 4a is shown a
plot of the sub-system dimension $d^{(s)}_L$ as a function of its size
$n_s$. A clear linear scaling relationship in the sub-system dimension is
seen with its size, indicating that analysis of relatively small sized
sub-systems may suffice in estimating the effective dimensionality of the
large system. It may be noted that similar interesting relationships in
Lyapunov dimensions as a function of lattice system size have been
observed in coupled maps [18] indicating a possible generalization.
Further, for analyzing the spatial complexities in the system dynamics we
studied the behavior of sub-system dimension density function
$\rho^{(s)}$ defined as
\begin{equation}
\rho^{(s)} = \lim_{ {n_s} \rightarrow \infty} {d^{(s)}_L \over n_s} .
\end{equation}\noindent 
A plot of $\rho^{(s)}$ as a function of sub-system size $n_s$ in fig. 4b
depicts a rapid convergence to a constant value ( $\sim 0.57$ for $n_s >
n_{sc}$) although the entropy (defined as the sum of the sub-system
positive Lyapunov exponents) increases linearly with $n_s$ (fig. 4c). This
increase in entropy suggests that the rate of information production and
the growth of uncertainity varies with the sub-system size. Further, the
converging behavior of $\rho^{(s)}$ may help in assessing the complexity
in the system dynamics. On comparing the magnitudes of $\rho^{(s)}$ for
the case without diffusion to that when diffusion is present, we find that
there is marked drop in its value from $2.01$ (without diffusion) to
$0.57$ (with diffusion). This implies that diffusion results in a decrease
in the average contribution to the Lyapunov dimension from each cell and
attempts to bring about spatial uniformity in the system variables. The
sensitive nonlinear features however play an antagonistic role. In
summary, a knowledge of the Lyapunov dimension $d^{(s)}_L$ and the
corresponding dimension density $\rho^{(s)}$ of a sub-system can help in
resolving the extents of complexity in the system dynamics.

We also studied the effect of varying the diffusion coefficients of the
interacting species (but maintaining their ratio constant) on the system
dynamics. In fig. 5 are shown plots of $D_1$ as a function of dimension
density $\rho^{(s)}$ (for $n_s = 7$) for two different ratios of
$D_3/D_1$. The figure shows the region where the spatio-temporal behavior
is more sensitive to the values of the diffusion coefficients and points
out that the chaotic patterns may break into coherent ones (at $D_1 =
0.75$). It may be clarified that similar behavior may also be observed for
variations in the other system parameters.

The characterization studies of coherent but self-replicating patterns
shown in fig. 2 yield a negative maximum Lyapunov exponent ($\lambda_{max}
\sim - 0.07$, calculated for $n_s = 21$). Thus, although the dynamics of
these patterns are complex, the negative value suggests that the pattern
would finally evolve to a stationary system solution.

\section{Synchronization of Spatio-Temporal Dynamics}

In this section, we shall analyze the capability of a suitably chosen
response system to synchronize its dynamics with that of a spatio-temporal
chaotic system using limited time-series data. In the context of
low-dimensional chaotic dynamics it has been observed that synchronization
is possible if conditional Lyapunov exponents turn out to be negative
[34,35]. The results obtained in Section 3 suggest that it may be
worthwhile to study whether it is sufficient to calculate the conditional
Lyapunov exponents for the sub-system of the response model to determine
its synchronizing ability.  Note that the response model has a reduced
dimensionality because it is driven by space-time signals in one of the
dependent variables $X_i$, $i = 1, 2,$ or $3$ . We shall present the
results on assuming that space-time data in $X_3 (j,t)$ are available. The
model of the response system then assumes the form
\begin{eqnarray}
{\partial \hat X_1(j,t)\over \partial t} = & 1 & - \hat X_1(j,t) 
- Da_1 \hat X_1(j,t) X_3^2(j,t) \nonumber \\
& + & D_1 [\hat X_1(j+1,t) - 2 \hat X_1(j,t) + \hat X_1(j-1,t)] \nonumber \\
{\partial \hat X_2(j,t)\over \partial t} = & \beta & - \hat X_2(j,t) 
- Da_2 \hat X_2(j,t) X_3^2(j,t) \nonumber \\
& + & D_2 [\hat X_2(j+1,t) - 2 \hat X_2(j,t) + \hat X_2(j-1,t)] ,
\end{eqnarray}\noindent
where $\hat X_1 (j,t)$ and $\hat X_2 (j,t)$, $j = 1,2, \dots ,N$ are the
response variables.

Similar to the sub-system Lyapunov exponents, it is possible to calculate
the conditional sub-system exponents by monitoring the growth rate of now
$2n_s$ sets of orthonormal vectors $\delta X_i (k,t)$, $i = 1, 2$, $k =
1,2,\dots n_s$ in a linearized region of (12) but in the reduced
dimensionality index $i$.

We found the maximum conditional sub-system exponent (for the chaotic
dynamics in fig. 1 with $n_s = 11$) to be negative indicating that the
response system may have the ability to synchronize. This was tested by
dynamically passing space-time signals in the variables $X_3 (j,t)$ ($j =
1, \dots , N$) from (3) to the response system defined by (12) (for $N =
64$). For the study it was further assumed that the response system did
not experience the finite amplitude perturbation and was only temporally
chaotic, while, (3) had evolved considerably into spatio-temporal chaos
before time-series signals in $X_3 (j,t)$ were passed to (12). The
simulation showed that the dynamics of the response system did synchronize
completely with the main system. This is depicted in figs. 6a and 6b where
the space-time errors converge to zero over the entire spatial domain in
both the dependent variables $e_i (t) = X_i (j,t) - \hat X_i (j,t)$,
($i=1,2$).

For the self-replicating patterns (fig. 2), the maximum sub-system
exponent calculated from (3) was found to be negative as reported above.
The evaluation of the conditional exponents is trivial and synchronization
is again likely to occur. From a different perspective, we have in an
alternate study, focussed our attention on the synchronization behavior of
these patterns with a view to controlling the dynamics of this interesting
class of spatio-temporal patterns [36].

\section{Conclusions}

In summary, the characterization of a reaction-diffusion system with
nonlinear autocatalytic kinetics and exhibiting complex dynamics has been
studied from a viewpoint of relating spatial sub-system dynamical behavior
to that of the system considered as a whole. The analysis indicates that
for the system exhibiting spatio-temporal chaos there exists a linear
scaling relationship in the Lyapunov dimension $d^{(s)}_L$ as the
sub-system size $n_s$ increases. For a set of operating parameter values
it is also seen that above a critical sub-system size $n_{sc}$ , the
Lyapunov density function $\rho^{(s)} (= d^{(s)}_L/n_s)$, which quantifies
the average information, saturates to a constant value. The growth of
uncertainity of the system and rate of information production evaluated in
terms of the K-S entropy, however, increases linearly with sub-system
size. These results may help in formulating strategies for embedding
complex dynamics of reaction-diffusion systems from time-series data.

Apart from chaotic spatio-temporal patterns this system is also seen to
exhibit other interesting patterns for suitable choices of parameter
values.  One such interesting pattern, viz., the self-replicating
patterns, has been demonstrated. The sensitivity of the complex dynamics
to operating values and especially the magnitudes of the diffusion
coefficient has been brought out.  Finally, the synchronization ability of
the dynamics of a response system using time-series data from the spatial
domain was analyzed for both spatio-temporally chaotic and
self-replicating dynamics. Our analysis indicates that if the maximum
conditional Lyapunov exponent of a sub-system is negative, synchronization
over the entire spatial domain is likely to occur.

{\bf Acknowledgements}
We gratefully acknowledge the support of the Department of Science and
Technology, New Delhi, India, in carrying out this work.

\pagebreak

\pagebreak
\begin{figure}
\caption{
\label{fig1}}
Spatio-temporal chaos arising for a perturbation given at $T = 40$. $D_1
= D_2 = 1.0$; $D_3 = 0.01$; $\alpha = 1.5$; $\beta = 2.93$; $Da_1 =
18000$; $Da_2 = 400$; $Da_3 = 80$; [ $X_1$ axis-scale: $(0.0,0.07)$; $x$
axis-scale: $(0.0,12.8)$; $t$ axis-scale: $(0.0,100.0)$ ].
\end{figure}
\begin{figure}
\caption{
\label{fig2}}
Self-replicating pattern for the three-variable autocatalytic reaction
$D_1 = D_2 = 1.0$; $D_3 = 0.01$; $\alpha = 1.0$; $\beta = 1.0$; $Da_1 =
50$; $Da_2 = 50$; $Da_3 = 2.95$; For the entire system (with $N = 256$)
placed in a homogeneous stationary state [$X_1 (0,0) = 1.0$, $X_2 (0,0) =
1.0$, $X_3 (x,0) = 0.0$] at $t = 0$, a finite perturbation $X_1 (0,0) =
0.5$, $X_2 (0,0) = 0.5$, $X_3 (x,0) = 0.25$ was given in the central 21
sites. [ $X_3$ axis-scale: $(0.0,2.96)$; $x$ axis-scale: $(0.0,51.2)$ ;
$t$ axis-scale: $(0.0,200.0)$ ].
\end{figure}
\begin{figure}
\caption{
\label{fig3}}
Sub-system dynamics for spatio-temporal chaos (a) $n_s = 7$; (b) $n_s =
21$. (c,d) The respective convergence of the maximum Lyapunov exponent for
the above sub-systems. System parameter values and other conditions of
analysis identical to fig. 1.
\end{figure}
\begin{figure}
\caption{
\label{fig4}}
Behavior of sub-system a) Lyapunov dimension, b) dimension density, and c)
entropy as a function of $n_s$ .
\end{figure}
\begin{figure}
\caption{
\label{fig5}}
Sensitivity of dimension density $\rho^{(s)}$ for varying $D_1$ for two
different ratios of $D_3/D_1$ ($D_2 = D_1$). Solid curve: $D_3/D_1 =
0.01$; dashed curve: $D_3/D_1 = 0.1$.
\end{figure}
\begin{figure}
\caption{
\label{fig6}}
Space-time errors converging to zero in both the dependent variables, $e_i
(t) = X_i (j,t) - \hat X_i (j,t)$, $j = 1,\dots,N$ (a) $i = 1$, (b) $i =
2$ indicative of complete synchronization in the spatio-temporal chaotic
dynamics of the system and its response model. System parameter values and
other conditions of analysis identical to fig. 1. Note the response model
had initial conditions (corresponding to temporal chaos) while the system
had developed into spatio-temporal chaos (i.e. 100 time units in Fig. 1)
at $t = 0$ when driving was switched on. $X_1$ axis-scale: (a) $(-24,26)$,
(b) $(-45,9.5)$; $x$ axis-scale: $(0.0,51.2)$; $t$ axis-scale:
$(0.0,200.0)$.
\end{figure}

\end{document}